\documentstyle[aps,psfig,multicol,prl]{revtex} 

\begin{document}

\draft

\twocolumn[\hsize\textwidth\columnwidth\hsize\csname @twocolumnfalse\endcsname

\title{Infrared Behavior of Interacting Bosons at Zero Temperature}

\author{ 
C.\ Castellani,$^1$ 
C.\ Di Castro,$^1$ 
F.\ Pistolesi,$^{2,3}$ 
and G.C.\ Strinati$^3$
} 

\address{
$^1$Dipartimento di Fisica, Universit\`a ``La Sapienza'', 
Sezione INFM, I-00185 Roma, Italy\\
$^2$Scuola Normale Superiore, 
Sezione INFM, I-56126 Pisa, Italy\\
$^3$Dipartimento di Matematica e
  Fisica, Universit\`a di Camerino, Sezione INFM, 
I-62032 Camerino, Italy
} 


\maketitle

\begin{abstract}
We exploit the symmetries associated with the stability 
of the superfluid phase to solve the long-standing problem 
of interacting bosons in the 
presence of a condensate at zero temperature. 
Implementation of these symmetries 
poses strong conditions 
on the renormalizations 
that heal the singularities of  
perturbation theory.
The renormalized theory gives:
For $d>3$ the Bogoliubov quasiparticles  as an exact result; 
for $1<d \le 3$ a nontrivial solution with the exact exponent
for the singular longitudinal correlation function,
with phonons again as low-lying excitations.
\end{abstract}
  
\pacs{Phys. Rev. Lett. {\bf 78}, 1612 (1997) \hfill cond-mat/9604076\hfill 
e-mail: {\em pistoles@ill.fr}}

\newpage
\vskip1pc]

The  problem of understanding the low-lying excitations from 
the ground state of an interacting Bose system has been one 
of the major problems of condensed matter theory in the 
fifties and sixties. 
The first solution to this problem was given by Bogoliubov 
as a generalized Hartree-Fock approximation\cite{B47}.
Numerous papers were then devoted to analyze the corrections 
to this approximate solution
\cite{Beliaev58,GavoretNozieres,DeP,Nepomshi,PS,Popov,Weichman88,NP,Stringari92,GB,Benfatto}.
Apart from approximations showing a gap in the excitation spectrum,
all attempts to improve the Bogoliubov approximation encountered 
the difficulty of a singular perturbation theory (PT) plagued by 
infrared (IR) divergences, due to the presence of the Bose-Einstein 
condensate and the  Goldstone mode
\cite{Beliaev58,GavoretNozieres}.

A systematic study of these IR divergences has been long delayed 
because they appeared only at intermediate steps
of the calculations while physical quantities turned 
out to be finite \cite{Beliaev58}.
That problems could  arise in PT
was originally recognised by Gavoret and Nozi\`eres 
\cite{GavoretNozieres}; indeed later \cite{Nepomshi}
it was found that the exact anomalous self-energy 
(at zero external momentum) has to vanish \cite{notaspin},
in contrast with 
the finite value obtained by the Bogoliubov approximation.
This result questions the validity of  
PT 
and requires a properly renormalized theory.

To take care of the IR  divergences 
at arbitrary spatial dimension $d$ greater than one, 
we exploit 
the renormalization-group (RG) approach.
In its standard application the RG approach sums up 
the singularities of PT and provides 
the power-law behavior of physical quantities which 
is characteristic of critical phenomena. Here 
we deal instead with a stable superfluid phase,
for which exact cancellations (insead of resummations)
of singular terms are expected to occur in physical response 
functions. It appears thus crucial to exploit the underlying
 (local-gauge) symmetry and the related Ward identitied (WI)
which implement these exact cancellations, as required on 
physical grounds. In this paper we combine the 
RG approach with the WI to obtain the solution 
to the problem \cite{comBen}.

To be more explicit we use the WI to: ({\em i})
Establish constraints on the RG procedure; 
({\em ii}) Relate renormalization parameters to physical 
quantities; ({\em iii}) Achieve the cancellation of 
singularities in the response functions \cite{MetDic}. 
In this way, the number of marginal and relevant 
running couplings (which are {\em a priori} necessary 
to study the IR behavior of interacting bosons
for $d\le 3$) is reduced to only one, {\em e.g.},
the longitudinal two-point vertex function $\Gamma_{ll}$.
In addition, we are able to close the equation for $\Gamma_{ll}$,
thus providing the {\em exact } IR behavior for the 
zero-temperature interacting Bose system. The resulting 
solution is quite different from the Bogoliubov one 
in spite of the coincidence of the linear spectrum.
In particular, 
we free the 
Gavoret and Nozi\`eres results from 
the provisions posed by the occurrence of 
IR divergences \cite{GavoretNozieres},
and recover the result of 
Ref.\ \cite{Nepomshi} for the anomalous self-energy.

We consider the following action for the Bose system:
\begin{eqnarray}
	S &=& 
	\!\!\int_0^\beta\!\!\!d\tau \!\! \int\!d {\bf x}
	\left\{ \vphantom{v\over2}
		\psi^*(x)
		\left[
			-\partial_\tau 
			+ \mu	
		\right]
		\psi(x)
		-\left|
			\left({\bf \nabla}-i {\bf A}\right)
		\psi(x) \right|^2
	\right.
	\nonumber \\
	&& 
	\left.
		-{v\over 2} 
		|\psi(x)|^4
		+\psi(x) \lambda^*(x)
		+\psi^*(x) \lambda(x)
		\vphantom{v\over2}
	\right\}\ ,
	\label{e1}
\end{eqnarray}
where $\psi(x)$  (with $x=(\tau,{\bf x})$) 
is a bosonic field obeying 
periodic boundary conditions in the imaginary time $\tau$ 
and $\beta$ is the inverse temperature 
(we set $\hbar=1$ and $m=1/2$).
In (\ref{e1})  $\lambda(x)$ and 
($\mu(x)$, ${\bf A}(x)$)$=A_\nu(x)$ 
($\nu= 0, \dots,\ d$) 
are external ``sources''
introduced to obtain the correlation functions.
At the end of the calculation 
$\mu(x)$ recovers the constant value $\mu$ 
of the  chemical potential while $\lambda$ and ${\bf A}$ 
are let to vanish.
 The interaction potential $v$ is taken to be  constant 
in  momentum space, as its  momentum dependence
is found to be irrelevant for the IR behavior.

The action (\ref{e1}) allows for spontaneous broken 
gauge symmetry; in that case, it is convenient to 
distinguish between longitudinal ($\psi_l$) and 
transverse ($\psi_t$) components to the 
broken-symmetry direction, by setting
$\psi(x) = \psi_l(x)+ i\psi_t(x)$ 
and 
$\psi^*(x) = \psi_l(x)- i\psi_t(x)$ (with a real order parameter).

By differentiating the free energy 
$F[A_\nu,\lambda_i] = \ln \int\!{\cal D}\psi_l {\cal D}\psi_t
\exp\left\{S\right\}$,
with $i=l,\ t$ and $\lambda 
= \lambda_l + i \lambda_t$, we obtain as
usual the connected correlation functions,
like the ``condensate wave function'' 
$
	\psi_{io} 
	= 
	\left\langle \psi_i(x) \right\rangle   
	=
	\delta F /\delta\lambda_i(x)
$
and the one-particle Green's function 
${\cal G}_{ij} = \delta^2F/\delta\lambda_i \delta \lambda_j$. 
It is further convenient 
to introduce the Legendre transform of $F$
with respect to $\lambda$, 
$\Gamma[A_\nu, \psi_{io}] 
= \int dx \lambda_i(x) \psi_{io}(x) -F[A_\nu,
\lambda_i]$,
whose derivatives are the vertex functions
$
	\Gamma_{{i_1\dots i_n};\nu_1\dots \nu_m}
	=
	{\delta^{(n+m)} \Gamma 
	/ 
	\delta \psi_{i_1 o} \dots \delta \psi_{i_n o}
	\delta A_{\nu_1} \dots \delta A_{\nu_m}
	} 
$
associated with 
the one-particle irreducible 
diagrams of PT.

In the  broken-symmetry phase we keep the value of the condensate  
$	
	\left\langle \psi_l(x) \right\rangle_{\lambda=0} 
	= 
	\psi_{lo}
$  fixed.
Accordingly, we introduce the fields $\tilde\psi_i$ with 
vanishing averages for vanishing external sources, such that 
$	\psi_l(x) = \psi_{lo}+ \tilde\psi_l(x)$ and  
$	\psi_t(x) = \tilde\psi_t(x)$.
The mean-field propagators ${\cal G}_{ij}$ are
 then obtained from the
quadratic part of the action 
$
	S_Q 
	= 
	-(1/2)\sum_k \tilde\psi_i(-k) 
	Q_{ij}(k) \tilde\psi_j(k)
$,
where
\begin{equation}
	Q(k) = 2 \left( 
		\begin{array}{cc}
		3v \psi_{lo}^2-\mu +{\bf k}^2 & -\omega_n \\
		\omega_n & v \psi_{lo}^2-\mu +{\bf k}^2
		\end{array}
		\right) 
	\label{e6}
\end{equation}
with $k=(i \omega_n,{\bf k})$ 
($\omega_n$ being a Matsubara frequency). 
In the following we consider the zero-temperature limit
where $\omega_n$ becomes a continuous variable $\omega$. 
Enforcing the
mean-field Bogoliubov  condition 
$\psi_{lo}^2 = \mu/v$ yields 
the IR behavior
${\cal G}_{tt}\sim (\omega^2+c_0^2{\bf k}^2)^{-1}$,
${\cal G}_{lt}\sim \omega(\omega^2+c_0^2{\bf k}^2)^{-1}$,
 and ${\cal G}_{ll}\sim {\bf k}^2(\omega^2+c_0^2 {\bf k}^2)^{-1}$,
where $c_0=\sqrt{2\mu}$ 
is the mean-field value of the sound velocity.
This singular IR behavior is associated with 
the presence of the Goldstone mode.
Recall that in the standard $\psi$-representation 
the Bogoliubov propagators 
(${\cal G}_{11}(k) = {\cal G}_{22}(-k)  
= {\cal G}_{tt}(k)+{\cal G}_{ll}(k)-2i{\cal G}_{lt}(k)$ and 
 ${\cal G}_{12}(k) = {\cal G}_{21}(k)=-{\cal G}_{tt}(k)+{\cal G}_{ll}(k)$) 
share 
a {\it common} $(\omega^2 +c_0^2 {\bf k}^2)^{-1}$ 
IR behavior, and the 
normal 
($\Sigma_{11}(k)=2\mu$)  
and the anomalous ($\Sigma_{12}(k)=\mu$) 
self-energies satisfy
 the Hugenholtz-Pines (HP) identity 
$\Sigma_{11}(0)-\Sigma_{12}(0) = \mu$ \cite{HP}.
In the $\psi_{l,t}$-representation,
on the other hand,  the 
various propagators have {\em different\/} 
IR behavior since
the Goldstone-mode singularity 
is kept in the transverse 
direction. This choice is crucial to select the
interaction terms according 
to their relevance \cite{Benfatto}.

To allow for the RG treatment,
it is convenient to rewrite the matrix $Q(k)$ in the more
general form   
\begin{equation}
	Q(k) = \left(
	\begin{array}{cc}
		v_{ll} + z_{ll}{\bf k}^2 
		+ u_{ll}\omega^2  
		& v_{lt} + w_{lt} \omega
		\\
		v_{lt} - w_{lt}\omega 
		& v_{tt} +z_{tt} {\bf k}^2 
		+ u_{tt}\omega^2
 	\end{array}
	\right)
	\label{e7}
\end{equation}%
where additional terms (running couplings)
have been introduced with respect to (\ref{e6}).
We also introduce running couplings for
cubic ($v_{ttt}$,  $v_{ltt}$, $\dots$) and 
quartic ($v_{tttt}$, $v_{lttt}$,  $v_{lltt}$, $\dots$) 
interaction terms, where the cubic terms originate 
from the presence of the condensate.
A perturbative expansion 
is then set up, as usual, by regarding the
quadratic action associated with (\ref{e7}) 
as the free 
part and the remaining terms as
perturbations.
In the absence of external sources $v_{tt}$ and $v_{lt}$
vanish by symmetry, as shown below.
The resulting  PT,
being massless,
 is plagued by IR divergences 
already at the one-loop level 
in spatial dimension $d \le 3$ 
\cite{Beliaev58,GavoretNozieres}.

Renormalization of the IR divergencies 
requires a preliminary power
counting for the running couplings. 
This is conveniently done by
keeping 
dimensionless the 
minimal set of couplings ($v_{ll}$, 
$w_{lt}$, $z_{tt}$) that yields the
linear part of the Bogoliubov spectrum.
We thus rescale them and the fields by 
 appropriate powers of $c_0$.
In this way, 
$[{\cal G}_{tt}]=-2$, $[{\cal G}_{lt}]=-1$,
and $[{\cal G}_{ll}]=0$,  
where $[A]$ stands for the engineering 
dimensions of $A$.
For simplicity,
from now on we shall omit indicating  $c_0$ 
whenever not strictly necessary.
We thus have $[{\bf x}] =-1$, $[\tau]=-1$,
$[\psi_l(x)] = (d+1)/2$,  $[\psi_t(x)] = (d-1)/2$, 
and 
$ 
	\left[{\Gamma}^{(n_l+n_t)}_{
	{l\dots l}{t\dots t}
	}(k_1, \dots )\right] 
	=  
	-n_l(d+1)/2-n_t (d-1)/2+d+1
$.
The upper critical dimension is $d_c=3$\cite{epsilon>2}.
For $d \leq 3$ the running couplings controlling the 
IR behavior have dimensions:
$	\left[v_{ll}\right] =  
	\left[w_{lt}\right] =  
	\left[u_{tt}\right] =  
	\left[z_{tt}\right] = 0 
$, 
$	
	\left[v_{ltt}\right] = \epsilon/2
$, 
 and 
$
	\left[v_{tttt}\right] = \epsilon
$, 
with $\epsilon = 3-d$. 
Although $v_{lt}$, $v_{tt}$, and $v_{ttt}$
would be strongly relevant, they vanish 
identically for vanishing
external sources.

One could proceed 
 at this point and derive the RG equations for the
running couplings. 
As mentioned above, however, 
contrary to critical phenomena in the 
present case of a stable phase, 
a singular PT has to 
result into  finite 
response functions. 
It is clear that cancellations of 
IR divergences in physical quantities
signal definite connections among the running couplings.
In the present context these
connections stem from the local 
gauge symmetry and are obtained by
examining the associated 
Ward identities \cite{MetDic}.

In our formalism the WI
result  from  the local gauge invariance of the
functional  $\Gamma$, namely, 
\begin{equation}
	\Gamma[A_\nu+\partial_\nu \alpha(x),
	 R_{ij}(\alpha(x)) \psi_{jo}] 
	= 	 
	\Gamma[A_\nu,\psi_{io}] 
	\label{e9}
\end{equation}
 $R_{ij}(\alpha)$ being the two-dimensional 
rotation matrix with 
 angle $\alpha$ in the space of the 
fields $\psi_l$ and $\psi_t$. 
This equation follows from the invariance of the action (\ref{e1})
under the gauge transformation 
$\psi(x) \rightarrow {\rm e}^{i \alpha(x)} \psi(x)$,
$\lambda(x) \rightarrow {\rm e}^{i \alpha(x)} \lambda(x)$,
and $A_\nu(x) \rightarrow A_\nu(x)+\partial_\nu\alpha(x)$
with $\alpha(x)$ real function.
Taking successive functional
 derivatives of (\ref{e9}) with respect
to $\alpha$, $\psi_{oi}$, and $A_\nu$ 
yields an infinite set of WI.
For our purposes only the following five WI are relevant.
The first two, which encompass the HP identity, 
are: 
\begin{eqnarray}
	&&\Gamma_{tl}(k) \psi_{lo} 
	+ \Gamma_t(0) - i k_\nu \Gamma_{l;\nu}(-k) = 0\,,
	\label{e10}
\\
	&&\Gamma_{tt}(k)\psi_{lo} 
	- \Gamma_l(0) 
	- i k_\nu \Gamma_{t;\nu}(-k) = 0\,.
	\label{e11}
\end{eqnarray}
In the limit $k_\nu\rightarrow 0$ they relate the two-point 
vertices to the external sources $\lambda_i = \Gamma_i(0)$,
and state that 
$v_{lt} = \Gamma_{lt}(0)$
and $v_{tt}=\Gamma_{tt}(0)$
 vanish when $\lambda_i=0$.
No gap thus appears in the one-particle spectrum. 
The second couple of WI 
\begin{eqnarray}
	&&\Gamma_{ttl}(k_1,k_2)\psi_{lo}
	+\Gamma_{tt}(-k_2) -\Gamma_{ll}(k_1+k_2) 
	\nonumber 
	\\
	&&\qquad
	- i \left(k_1\right)_\nu 
	\Gamma_{tl;\nu}(k_2,-k_1-k_2) = 0\,,
	\label{e12}
\\
	&&
	\Gamma_{ttt}(k_1,k_2)\psi_{lo}
	-\Gamma_{lt}(-k_2) -\Gamma_{lt}(k_1+k_2) 
	\nonumber 
	\\
	&&\qquad
	- i \left(k_1\right)_\nu 
	\Gamma_{tt;\nu}(k_2,-k_1-k_2) = 0\,,
	\label{e12.1}
\end{eqnarray}
are the standard WI associated with the continuity 
equation, modified now by the presence of the 
three-point vertices. In the limit $k_1=0$ and $k_2\rightarrow 0$
they yield $v_{ltt} \psi_{lo} = v_{ll}$ and $v_{ttt}= 0$.
The last of our WI is 
\begin{eqnarray}
	&&	
	{\Gamma_{tttt}}(k_1,k_2,k_3)\psi_{lo}
	\nonumber \\ 
	&&
		-{\Gamma_{ltt}}(-k_2-k_3,k_2) 
		-\!{\Gamma_{ltt}}(k_1+k_3,k_2) 
		-\!{\Gamma_{ltt}}(k_1+k_2,k_3) 
	\nonumber \\
	&& 
	- i\left(k_1\right)_\nu 
	{ \Gamma_{ttt;\nu}}(k_2,k_3,-k_1-k_2-k_3) = 0\,, 	
	\label{e13}
\end{eqnarray}
from which we obtain 
$
	v_{tttt}\psi_{lo}/3 =
	v_{ltt}
	\label{enuova}
$
for vanishing $k$'s.

From the above WI we also relate the running couplings to 
the (composite) current vertices and response functions.
Specifically, in the limit $k\rightarrow 0$ we obtain:  
\begin{equation}
	w_{lt} =  
		{1\over \psi_{lo}} 
		{\partial^2 \Gamma 
		\over 
		\partial \psi_{lo} \partial \mu},
	\quad 
	u_{tt} =
		-{1\over \psi_{lo}^2} 
		\left.{\partial^2 \Gamma \over 
		\partial \mu^2}\right|_{\psi_{lo}},
	\quad
	z_{tt} = 
	{2 n_s\over \psi_{lo}^2}, 
	\label{e15} 
\end{equation}
where $n_s$ is the superfluid density
(Josephson identity). We also have 
$
	v_{ll} 
	= 
	{(\partial^2 \Gamma /\partial \psi_{lo}^2})_\mu 
$
 by its very definition \cite{similar}.
We are left eventually 
with four running couplings, namely,
 $v_{ll}$, $w_{lt}$, $u_{tt}$, and $z_{tt}$, 
whose IR behavior can be obtained exactly.

As a guide to the procedure 
for obtaining this behavior, 
the  RG  equations will be evaluated 
at the one-loop level. 
Exploiting the $\epsilon$-expansion and the 
minimal subtraction technique we get:
\begin{equation}
\begin{array}{rclrcl}
	\displaystyle	
	s {dv_{ll} \over d s} 
	&=& \vspace{.3cm}
	\displaystyle 
	{c\, v_{ll}^2 
	\over 
	2 {\bar \psi}_{lo}^2\, z_{tt}^2}\,,
	\qquad
	& 
	\displaystyle	s {dw_{lt} \over d s} 
	&=&
	\displaystyle 
	{c\,v_{ll}\, w_{lt} 
		\over 
	2 {\bar \psi}_{lo}^2\, z_{tt}^2}\,,
	\\
	\displaystyle	s {du_{tt}\over d s} 
	&=& 
	\displaystyle
	-{c\, w_{lt}^2 
		\over 
	2 {\bar \psi}_{lo}^2\, z_{tt}^2}\,,
	& 
	\displaystyle	s {dz_{tt}\over d s} 
	&=&
	0\,,
\end{array}
\label{e16}
\end{equation}
where $s=\kappa'/\kappa$ is the scaling factor
($\kappa$ being the normalization point), 
$
	c(s)^2 
	= 
	v_{ll}(s) z_{tt}(s) /(
	v_{ll}(s) u_{tt}(s) + w_{lt}(s)^2)	 
$
is the square velocity entering  
the one-particle 
propagator according to (\ref{e7}),
and  
$\bar\psi_{lo}(s) 
= \psi_{lo} \kappa^{\epsilon/2} s^{\epsilon/2}$ 
\cite{notaKd}.
When $d=3$ Eqs.~(\ref{e16}) 
are equivalent to those of 
Ref.~\cite{Benfatto}, provided the 
coupling of Ref.~\cite{Benfatto} 
analogous to $v_{tttt}$  is 
identified with $3 v_{ll}/\psi_{lo}^2$,
consistently with the above results.

Quite generally, the solutions 
of the coupled equations
(\ref{e16}) take the form:
\begin{equation}
\begin{array}{rcl}
	z_{tt}(s) 
	&=& 
	z_{tt}(1)\,, 
	\qquad

	{ w_{lt}(s)} 
	= 
	\displaystyle	
	{w_{lt}(1) 
	\over v_{ll}(1) } { v_{ll}(s)} \,,
	\\ 
	u_{tt}(s)
	&=& 
	\displaystyle
	u_{tt}(1)+{w_{lt}(1)^2 \over v_{ll}(1)}
	-
	{w_{lt}(1)^2 \over v_{ll}(1)^2} 
	{v_{ll}(s)} \,. 
	\label{e17}
\end{array} 
\end{equation}
This implies that it is sufficient to 
determine $v_{ll}(s)$.

Although we have derived 
 (\ref{e17}) at the one-loop order,
we {\it expect\/} them to hold exactly 
on physical ground 
owing to  the identification (\ref{e15})
of the renormalization parameters 
with physical quantities.
To begin with, the $s\rightarrow 0$ value 
of $z_{tt}$ is the ratio $n_s/n_0$ of {\em finite} 
physical quantities ($n_0=\psi_{lo}^2$ being 
the condensate density)
so that divergences compensate each other
in its  expression, leading to the first of (\ref{e17}).
Regarding the second of (\ref{e17}), 
we obtain from (\ref{e15}) for $w_{lt}$ 
and the definition of $v_{ll}$ 
that the ratio
$v_{ll}/w_{lt}$ reduces to 
$-2n_0/(d n_0/d \mu)_\lambda$ 
in the limit $k\rightarrow0$.
Here $(d n_0/d \mu)_\lambda$ is   
the ``condensate compressibility''
which has to be finite for thermodynamic stability.
Finally, from the definition of $c(s)^2$ and 
 (\ref{e15}) 
we obtain  that $c(s)^2$ 
reduces to  $c^2$ in the limit $s\rightarrow 0$, 
where now $c^2=2n_s/(d n /d \mu)_\lambda$ 
is the square of the macroscopic sound velocity 
($n$ being the density and $n=n_s$ 
at zero temperature).
By the very stability of the bosonic system, 
$c^2$ is free from IR divergences,
thus suggesting that $c(s)$ is finite and does not scale with $s$, 
{\em i.e.}, $c(s)=c=c_0$, 
apart from finite corrections 
originating  from nonsingular terms that do 
not enter the RG flow.
Exploiting the first and second of 
(\ref{e17}), we verify that 
the condition  $c(s)=$ constant
reduces to the third of (\ref{e17}).

The {\it proof\/} of the above statements 
is as follows.
$z_{tt}$ has to remain constant by inspection of
the WI (\ref{e11}), which shows that the  
divergence of $z_{tt}$
expected by power counting is actually not
present, 
since it is related to non diverging quantities. 
$w_{lt}$ can instead be identified with $v_{ll}$
via the WI (\ref{e12}) and (\ref{e12.1}), 
which relate 
$v_{ll}$ to $\Gamma_{ltt}$ and 
$w_{lt}$  to  $\Gamma_{tt;0}$, respectively,
the latter identification being obtained from the $\omega$-derivative 
of (\ref{e12.1}).
By inspection of the leading singular terms to all 
orders in PT, 
$\Gamma_{tt;0}$
and $\Gamma_{ltt}$ are then found to be proportional 
to each other \cite{nostro};
by the same procedure,  the invariance 
of $c(s)$ implied by the last 
of (\ref{e17}) follows from 
the exact connection between the 
singular parts of 
$\Gamma_{;00}$ 
 and 
$\Gamma_{ll}$,
associated  respectively with $u_{tt}$ and $v_{ll}$.

Determining the IR behavior 
is thus exactly reduced to solving for a {\it single} running
coupling, for instance $v_{ll}$. 
In particular, at the one-loop order we obtain:
\begin{equation}
	{v_{ll}(1) \over v_{ll}(s)} 
	=
	\left\{
	\begin{array}{cc}
	\displaystyle		
	1 - 	{v_{ll}(1) \over 
			 2 \bar \psi_{lo}(1)^2 
			z_{tt}(1)^2 } 
			{\rm ln} s 
		&(\epsilon=0)\\
		\displaystyle		
	1 + 
		{
		v_{ll}(1) (s^{-\epsilon}-1)
		\over 
		2 \bar \psi_{lo}(1)^2 z_{tt}(1)^2 \epsilon
		} 
	&(0<\epsilon<2)\,.
	\end{array}
	\right.
\label{e20}
\end{equation}
In both cases $v_{ll} \rightarrow 0$ 
as $s\rightarrow 0$, while for 
$\epsilon<0$ $v_{ll}$ remains finite.
We show below that the asymptotic 
behavior (\ref{e20})
of $v_{ll}$  is actually exact.

The one-particle Green's 
function resulting from (\ref{e17})
and (\ref{e20}) have the form \cite{notaKd}:
\[
\begin{array}{rcl}
	{\cal G}_{ll}(k) 
	&=& 
	\left\{
	\begin{array}{lc}
		\displaystyle
		-{c\,n_0 \ln k 
		\over 64 \pi^2 n_s^2}
		& (\epsilon=0)
		\\
		\displaystyle
		{c\,n_0  K_{4-\epsilon} 
		 \over 
		8 \epsilon n_s^2}	
		k^{-\epsilon} 
		& (0<\epsilon<2)
	\end{array}
	\right\}
	\quad \sim \displaystyle	
	{1\over v_{ll}}, 
\\
	{\cal G}_{lt}(k) 
	&=& \displaystyle	
	{d n_0 \over d \mu} 
	{c^2\over 4 n_s} \,  
	{\omega \over \omega^2+c^2{\bf k}^2}
	\quad\sim  \displaystyle
	-{w_{lt} \over v_{ll}}{ 1 \over z_{tt}} \, 
	{\omega \over {\bf k}^2 + \omega^2/c^2},
	\\
	{\cal G}_{tt}(k) 
	&=& \displaystyle	{c^2 \ n_0 \over 2 n_s}
	{1\over \omega^2+c^2{\bf k}^2} 
	\quad \sim  
	\displaystyle	
	{1\over z_{tt}}\
  	{1\over {\bf k}^2 + \omega^2/c^2 }\,,
	\end{array}	
\]
where  the asymptotic 
($k\rightarrow 0$) values of the running
couplings have been identified 
via (\ref{e15}).
Note that the IR behavior 
of ${\cal G}_{lt}$ and ${\cal G}_{tt}$
is completely and exactly
 determined by the conditions (\ref{e17}), 
and is independent from $d$ and the behavior of $v_{ll}$.
Instead ${\cal G}_{ll}$ diverges 
logarithmically as $k\rightarrow 0$  when $d=3$ and 
like $k^{-\epsilon}$ for $1<d<3$ 
\cite{Nepomshi,PS,Weichman88}.
Accordingly, we find that the 
anomalous self-energy $\Sigma_{12}(k)$ vanishes 
like $1/\ln k$ or like $k^{\epsilon}$, 
with the  HP identity now reading 
$\Sigma_{11}(0)=\mu$ \cite{Nepomshi}.
In addition to recovering the 
result by Gavoret and Nozi\`eres 
\cite{GavoretNozieres} for the propagators  
in the $\psi$-representation,
we have also obtained the divergent
subleading {\em ln} terms \cite{logterm}.
Such {\em ln} terms, on the other hand, 
can be shown to disappear
altogether in the expressions for the density-density,
 density-current, and current-current 
response functions
\cite{Nepomshi,nostro}.

Although $v_{ll}(s)$ has been explicitly obtained 
at the one-loop order, we show now that its asymptotic 
behavior is exact. 
In fact, for $d>3$ all interactions are
irrelevant, the Bogoliubov result is 
correct, and no longitudinal
divergence appears. 
For $d=3$, all perturbation couplings to the
Bogoliubov  action are marginally 
irrelevant \cite{Benfatto},
 and the result obtained at the 
one-loop order is stable. 
For $1<d<3$, the one-loop 
calculation presented above
leads to  a nontrivial fixed point 
({\it i.e.}, 
$v_{tttt}(s)/\kappa^\epsilon \rightarrow v^* \neq 0$) 
with $v_{ll} \sim s^{\epsilon}$.
The WI result 
$v_{ll}=v_{tttt}  
\bar\psi_{lo}(1)^2 \kappa^{-\epsilon} s^\epsilon/3$
 guaranties that 
the exponent of the asymptotic behavior 
of $v_{ll}$ remains $\epsilon$, 
{\it irrespective} of the actual value $v^*$ of 
the nontrivial fixed point, 
provided it exists.
That $k^\epsilon$ is actually
the true asymptotic behavior 
of $\Gamma_{ll}(k)$ 
follows by expressing the
 singular part of $\Gamma_{ll}$ 
in terms of the exact 
$\Gamma_{ttl}$ and ${\cal G}_{tt}$: 
\begin{equation}
	\Gamma_{ll}(k) 
	= 
	B- {v_{ll}^o \over 2} 
	\sum_q {\cal G}_{tt}(q){\cal G}_{tt}(q+k)\Gamma_{ttl}(q, -q-k)
	\label{ee18}
\end{equation}
where $B$ coincides with the bare coupling 
$v_{ll}^o = v_{ll}(s=1)$ 
apart from finite contributions.
The WI (\ref{e12}) allows eventually to express 
$\Gamma_{ttl}$ in terms of 
$\Gamma_{ll}$ itself, thus 
closing Eq. (\ref{ee18}) and providing the 
exact asymptotic behavior $\Gamma_{ll}(k) \sim k^\epsilon$. 
An analogous procedure was implemented in Ref. \cite{Nepomshi}.

Some final comments are in order.
Contrary to the theory of critical phenomena, where 
the WI do not provide stringent conditions 
 owing to the genuine divergences
of physical quantities 
(such as the compressibility in the gas-liquid transition), 
in our case the relevant physical quantities are finite and
the WI imply 
cancellations in the corresponding couplings. 
In this way, lines of fixed points have to appear 
corresponding to different finite values of the 
thermodynamic derivatives. 
It is also worth noticing that the 
Bogoliubov mean-field  solution for $1< d \le3$ 
is infinitely distant  (in the RG sense) from our 
nontrivial fixed point 
whenever  the interparticle interaction is nonvanishing.
This is true despite the fact that 
the resulting physical picture 
contains the {\em same} low-lying elementary excitations.

One of us (F.P.) gratefully acknowledges 
partial financial support 
from Europa Metalli-LMI S.p.A. and from the 
Italian INFM through the
Unit\`a di Camerino.


\begin{references}
\bibitem{comBen}
	A similar RG approach was used in 
	Ref.\ \cite{Benfatto}, although in $d=3$ only
	and without a systematic use of WI; 
	WI have been partially exploited already in reference 	
	\cite{GavoretNozieres,DeP}.
\bibitem{epsilon>2} 
	For $d=1$ (and zero temperature) the Mermin-Wagner-Hohenberg
	theorem states that $\psi_{lo}=0$, thus defining the lower critical
	dimension.
\bibitem{notaKd}
	The factor $K_{d+1} = 2[(4\pi)^{(d+1)/2} 
	\Gamma((d+1)/2)]^{-1}$ 
	has been absorbed in the definitions 
	of $v_{ll}$  
	and $w_{lt}$, $\Gamma$ here being 
	the Euler's gamma function
\end{references}
\end{document}